\documentclass{article}
\usepackage[utf8]{inputenc}
\usepackage{natbib}
\usepackage{amssymb}
\usepackage{pifont}
\usepackage{bm}
\usepackage{mathalfa} 
\usepackage{amsmath}
\usepackage[dvipsnames]{xcolor}
\usepackage{comment}
\usepackage{soul}
\usepackage{multirow}
\usepackage{url}

\renewcommand{\baselinestretch}{1.5}
\newcommand {\bd}[1]{\mbox{\boldmath$#1$}}

\newcommand{\ind}{\perp\!\!\!\!\!\!\perp}

\title{Direct Estimation for Commonly Used Pattern-Mixture Models in Clinical Trials}
    \author{\\ \bf Jitong Lou, Mallikarjuna Rettiganti, Yongming Qu* \\\\
Department of Global Statistical Sciences \\ Eli Lilly and Company, Indianapolis, Indiana, USA \\ *Email: qu\_yongming@lilly.com \\\\}

\date{\today}

\usepackage{natbib}
\usepackage{graphicx}

\begin{document}

\maketitle

\begin{abstract}
Pattern-mixture models have received increasing attention as they are commonly used to assess treatment effects in primary or sensitivity analyses for clinical trials with nonignorable missing data. Pattern-mixture models have traditionally been implemented using multiple imputation, where the variance estimation may be a challenge because the Rubin’s approach of combining between- and within-imputation variance may not provide consistent variance estimation while bootstrap methods may be time-consuming. Direct likelihood-based approaches have been proposed in the literature and implemented for some pattern-mixture models, but the assumptions are sometimes restrictive and the theoretical framework is fragile. In this article, we propose an analytical framework for an efficient direct likelihood estimation method for commonly used pattern-mixture models corresponding to return-to-baseline, jump-to-reference, placebo washout, and retrieved dropout imputations. A parsimonious tipping point analysis is also discussed and implemented. Results from simulation studies demonstrate that the proposed methods provide consistent estimators. We further illustrate the utility of the proposed methods using data from a clinical trial evaluating a treatment for type 2 diabetes.
\end{abstract}

\section{Introduction}
Missing data are common in clinical trials and can be either a result of missed measurements or unobservable potential outcomes due to intercurrent events \citep{international2020harmonised}. With ignorable missingness, where the ultimate outcome is independent of the missingness indicator given the observed values, the maximum likelihood method can be used to estimate the treatment effect. However, in many cases, the ignorable missingness assumption may be violated or cannot be verified, so a nonignorable missingness assumption may be used for primary or sensitivity analyses. In addition, estimands may be defined by a hypothetical strategy conditioning on certain intercurrent events and the unobserved potential outcome \citep{lipkovich2020causal,qu2021implementation}. In these cases, the pattern-mixture model \citep{little1993pattern,little1994class} is an attractive approach to handling missing values that are either unobserved or unobservable (censored due to using a hypothetical strategy to handle intercurrent events). 

Three commonly used pattern-mixture models to handle missing values include using the pattern of (1) the baseline value of the outcome variable, called the return-to-baseline (R2B) imputation \citep{leurent2020reference, qu2022return}; (2) the reference group, including the jump-to-reference (J2R), copy-reference (CR), copy-incremental-reference, and placebo washout (PW) imputations \citep{carpenter2013analysis, wang2023statistical}; and (3) the retrieved dropouts, referred to as the retrieved dropouts (RD) imputation \citep{CHMP2010, mcevoy2016missing}. 
The RDs are those participants who discontinue study treatment but remain in the study. Consider the example of a clinical trial participant who discontinued the study medication due to an adverse event and subsequently received an alternative rescue medication. In this case, the potential outcome for this participant if they had continued without rescue medication was not observed, resulting in a missing value. This participant might be considered to have no benefit or a partial benefit from the study medication, as this participant was unable to tolerate the medication \citep{qu2021defining,qu2021implementation}. Depending on the situation, one of the above three pattern-mixture models could be used to handle this missing value. 

The three aforementioned pattern-mixture models are often implemented using multiple imputation \citep{rubin1976inference}. 
While multiple imputation is a flexible and convenient method to implement these pattern-mixture models, a few drawbacks exist. First, multiple imputation is a Monte Carlo-based method that may require a large number of imputations to achieve an accurate estimate. Second, the variance estimation based on the Rubin method of combining within- and between-imputation variance \citep{barnard1999miscellanea} is often invalid valid due to the uncongeniality between the imputation and analysis models \citep{meng1994multiple,xie2017dissecting,bartlett2020bootstrap}. {Although \cite{bartlett2020bootstrap} showed when bootstrap is used appropriately a consistent variance estimation can be obtained, the bootstrap procedure (embedded by multiple imputation) is time consuming}. Finally, if multiple analyses are needed for the same variable, the imputed data may need to be stored in a database, leading to a sizeable analysis dataset. If data are not stored, the imputation procedure needs to be repeated for every analysis, even for the same variable.

{These pattern-mixture models can be solved by a joint likelihood under the framework by \cite{little1993pattern}. However, maximizing the likelihood for each pattern mixture model requires customized programming, which may be a challenge for most clinical statisticians.} Efforts have been made to estimate some of the above pattern-mixture models {using the (restricted) maximum likelihood estimates under MAR, which can be readily produced by commonly used statistical software packages,} without multiple imputation. It is often called the \emph{direct likelihood} estimation approach, although the direct estimation does not need to be based on the maximum likelihood estimation. In this article, we refer to this approach as \emph{direct estimation}. \cite{mehrotra2017missing} provided a direct estimation method for a \emph{control-based mean imputation}, but the estimator is not equivalent to the J2R or CR imputation. \cite{zhang2020likelihood} provided a direct estimation method for the R2B pattern-mixture model, but the authors assumed the independence of the estimator for the means under the missing at random (MAR) assumption and the proportion of missing values, which is a strong assumption. More recently, \cite{garcia2023flexible} developed methods using analytical approach for estimating the treatment effect using various pattern-mixture models, including the J2R and CR patterns by modeling both the outcome variable and the intercurrent events. The use of direct estimation for the RD pattern has not been studied in literature. 

In this article, we {clearly formulated the potential outcome of the imputation, and} proposed direct estimation methods for the four commonly used pattern-mixture models {(R2B, J2R, PW, and RD) by leveraging the existing software packages for linear models}. In the new methods, we applied sandwich variance estimation and a novel method adjusting for baseline covariates. In addition, $\delta$-adjusted sensitivity analyses with an analytical approach to identify the tipping points are also presented. This article is arranged as follows. Section \ref{sec:methods} introduces the needed notations and describes the estimators for the four pattern-mixture models along with the variance estimation, the $\delta$-adjusted sensitivity analysis, and the methods for adjusting for baseline covariates. Section \ref{sec:simulation} compares the new direct estimation methods with the estimators based on multiple imputation and the existing direct estimation methods in the literature. In Section \ref{sec:application}, we apply the new methods to data from a clinical trial and interpret the results. Finally, in Section \ref{sec:summary} we summarize the methods and discuss its advantages and extended use. 

\section{Methods} \label{sec:methods}
Consider a clinical trial with $I+1$ parallel randomly assigned treatment groups indexed by $i=0,1,\ldots,I$ with $i=0$ representing the reference arm. Let $\bd Y_j=(Y_{j0}, Y_{j1}, \ldots, Y_{jK})'$ denote a vector of variables for the longitudinal outcome of interest for time  $0=t_0<t_1<\ldots<t_K$ for participant $j$ ($j$ = 1, 2, $\ldots$, $n$), $T_j=0, 1, \ldots, I$ denote the treatment group to which participant $j$ is assigned, and $\bd X_j$ denote a vector of $m$ baseline covariates. Generally, $Y_{j0}$ is the baseline value of the dependent variable and $Y_{jK}$ (at the final time point) is the primary endpoint for the analysis. Without loss of generality, we assume $\bd X_j$ includes $Y_{j0}$, the baseline value of the dependent variable. Since a mixed strategy may be commonly used to handle intercurrent events in defining estimands and missing values can occur even without intercurrent events, more than one pattern of imputation are often required in estimating the estimands \citep{international2020harmonised, darken2020attributable,meyer2020statistical,qu2021defining,qu2021implementation}. Therefore, we will consider a mixture of the following two patterns: ``adherent or could be adherent" (Pattern A) and ``could not be adherent" (Pattern B). Pattern A includes data for participants who are either adherent to the study treatment, could be adherent to study treatment in real life but not adherent in clinical trials due to reasons not related to the study treatment (e.g., pandemic, natural disaster, geographic conflict, scheduling conflict, travel, relocation, etc), {or experienced an intercurrent event that is handled by the hypothetical strategy}. Pattern B includes data for participants who are unable to adhere to study treatment due to insurmountable reasons such as tolerability and safety. {We assume the intercurrent events in Patter B are handled by the treatment policy strategy.} There may be many ways to define each pattern, and the statistical methods discussed in this article work regardless of the definition of the patterns. Let $A_{jk}$ be the pattern indicator at time point $k$ ($1 \le k \le K$) for participant $j$ with 
\begin{flalign}
\quad\quad A_{jk} = \left\{
\begin{array}{l}
1, \mbox{ Pattern A}  \\
0, \mbox{ Pattern B}
\end{array} \right. && \nonumber
\end{flalign}
and $R_{jk}$ be the indicator for missing data at time point $k$ ($1 \le k \le K$) for participant $j$ with
\begin{flalign}
\quad\quad R_{jk} =  \left\{
\begin{array}{l}
1, \mbox{ } Y_{jk} \mbox{ is observed}  \\
0, \mbox{ } Y_{jk} \mbox{ is missing}
\end{array} \right. . && \nonumber
\end{flalign}
{There are 3 possible scenarios that can cause missing values in Pattern A: i) participants can be adherent to study treatment but the measurement of the primary outcome could be missing or invalid. For example, a glycated hemoglobin (HbA1c) measurement is missing because the blood sample is compromised; ii) participants may have discontinued the study due to administrative reasons resulting in treatment discontinuation and missing values. For example, a participant relocates and has to discontinue the study and treatment altogether; and iii) participants have an intercurrent event that is handled by the hypothetical strategy. The outcome under the hypothetical strategy is unobservable and is considered missing in the analysis. For example, a participant may initiate an anti-diabetic concomitant medication due to high intermediate HbA1c measurements. It is reasonable to assume that the data are missing at random (MAR) for the missing values in each of the above three scenarios. 
} Subjects in Pattern B with $R_{jK} = 1$ are often referred as RDs. If there is a sufficient number of RDs, it is generally the preferred approach to use the RDs to impute the missing data in Pattern B. However, when the number of RDs is small, the R2B or reference arm-based imputation may be used. 
To simplify the notation throughout the rest of this article, we use $Y_j$ for $Y_{jK}$, $R_j$ for $R_{jK}$, and $A_j$ for $A_{jK}$ to denote the outcome, the indicator for missing data, and the pattern indicator, respectively. Similar notation is used for the missing indicator $R$ and the pattern indicator $A$. 

Using the potential outcome framework in defining estimands \citep{lipkovich2020causal}, let $Y_{jk}(i,b)$  denote the potential outcome at time point $k$ under treatment regimen $b$ when assigned to treatment group $i$. Under the treatment-regimen estimand, one may not care about $b$, so we use $Y_{ik}(i)$ to replace $Y_{jk}(i,b)$ for the potential outcome. Throughout the article, we use the following assumptions:
\begin{flalign}
\quad\quad \mbox{A1}:   \quad \quad  \left\{
\begin{array}{l}
Y_{jk} = \sum_{i=0}^I I(T_j=i) \cdot Y_{jk}(i)  \\
R_{jk} = \sum_{i=0}^I I(T_j=i) \cdot R_{jk}(i)  \\
A_{jk} = \sum_{i=0}^I I(T_j=i) \cdot A_{jk}(i) 
\end{array} \right. && \nonumber
\end{flalign}
and
\begin{flalign}
 \quad\quad \mbox{A2}: \quad\quad \{Y_{j1}(i), Y_{j2}(i), \ldots, Y_{jK}(i); R_{j1}(i), R_{j2}(i), \ldots, R_{jK}(i); A_{j1}(i), A_{j2}(i), \ldots, A_{jK}(i)\} \ind T_j . && \nonumber
\end{flalign}
Assumption A1 is called the stable unit treatment value assumption \citep{imbens2015causal}. 
Assumption A2 is the treatment ignorability assumption, which holds for randomized clinical trials. 

For convenience, we also assume the distribution of $\bd Y_j(i)$ conditional on $\bd X_j$, $\bd A_j(i)$, and/or $\bd R_j(i)$ is a normal distribution and that linear models can be used to estimate the relevant parameters, where $\bd A_j(i) = [A_{j1}(i), A_{j2}(i), \ldots, A_{jK}(i)]'$ and $\bd R_j(i) = [R_{j1}(i), R_{j2}(i), \ldots, R_{jK}(i)]'$.  We assume the baseline covariates $\bd X_j$ have no missing values.

\subsection{R2B Imputation} \label{sec:r2b}
For treatments with a potential acute effect to control for the symptoms or biomarkers of the disease without disease modification, it is reasonable to assume that participants who discontinue treatment would have no benefit from the treatment and that symptoms or biomarkers would return to baseline levels. As such, when there are not enough RDs, the R2B imputation can be used in which missing values may be imputed assuming they have the same mean as those at baseline. Under the assumption that the number of RDs is small, they can simply be grouped with Pattern A and thus it can be assumed that all participants without missing values belong to Pattern A and all patients in Pattern B have missing values. Thus, we have the following assumption:
\begin{flalign}
\label{eq:A_R}
\quad\quad & \mbox{A3}:   \quad \quad  R_{jk}(i) = 0 \mbox{ if } A_{jk}(i) = 0. && \nonumber
\end{flalign}
The potential outcome using the MAR imputation for missing values within Pattern A and the R2B imputation \citep{leurent2020reference,qu2022return} for missing values within Pattern B is given by
\begin{eqnarray}
Y_{jk}^{\mbox{\footnotesize MAR-R2B}}(i) =
A_{jk}(i) R_{jk}(i) Y_{jk}(i) + A_{jk}(i)\{1-R_{jk}(i)\} Y_{jk}^{\mbox{\footnotesize MAR}}(i) +  \{1-A_{jk}(i)\} \{1-R_{jk}(i)\} Y_{jk}^{\mbox{\footnotesize R2B}}(i),
\label{eq:RTB_PO}
\end{eqnarray}
where $Y_{jk}^{\mbox{\footnotesize MAR}}(i)$ is the potential outcome for the imputed value under the MAR assumption,
$Y_{jk}^{\mbox{\footnotesize R2B}}(i) = Y_{jk}^{\mbox{\footnotesize MAR}}(i) - \mu_{ik}^{\mbox{\footnotesize MAR}} + \mu^b$ is the potential outcome for the imputed value under the R2B imputation, $\mu_{ik}^{\mbox{\footnotesize MAR}} = E[R_{jk}(i) Y_{jk}(i) + \{1-R_{jk}(i)\} Y_{jk}^{\mbox{\footnotesize MAR}}(i)]$ the expected mean under the MAR assumption, and $\mu^b = E\{Y_{j0}(i)\}$. 

\cite{qu2022return} provided a convenient R2B imputation utilizing the existing imputation procedures under the MAR assumption as 
\[
Y_{jk}^{\mbox{\footnotesize R2B}, (r)}(i) = Y_{jk}^{\mbox{\footnotesize MAR},(r)}(i)-\hat \mu_{ik}^{(r)}+\hat \mu^b,  \; \; \mbox{ for } R_{ji}(i) = 0,
\]
where $Y_{jk}^{\mbox{\footnotesize MAR},(r)}(i)$ is the $r^{\mbox{th}}$ imputed sample under the MAR assumption, $\hat \mu_{ik}^{(r)}$ is the sample mean of $\left\{R_{jk}(i) Y_{jk}(i) + \{1-R_{jk}(i)\} Y_{jk}^{\mbox{\footnotesize MAR},(r)}(i), j=1,2,\ldots,n \right\}$ for treatment $i$ at time point $k$, and $\hat \mu^b$ is the sample mean of $Y_{j0}(i)$ across treatment groups.

Based on Assumption A3, the estimand corresponding to the potential outcome in Equation (\ref{eq:RTB_PO}) is
\begin{eqnarray}
\mu_{ik}^{\mbox{\footnotesize MAR-R2B}}  &=& E[A_{jk}(i) R_{jk}(i) Y_{jk}(i) + A_{jk}(i)\{1-R_{jk}(i)\}  Y_{jk}^{\mbox{\footnotesize MAR}}(i) \nonumber\\ && +  \{1-A_{jk}(i)\} \{1-R_{jk}(i)\}  \{Y_{jk}^{\mbox{\footnotesize MAR}}(i) - \mu_{ik}^{\mbox{\footnotesize MAR}} + \mu^b\}] \nonumber\\
&=& E[R_{jk}(i) Y_{jk}(i) + \{1-R_{jk}(i)\}  Y_{jk}^{\mbox{\footnotesize MAR}}(i)] - E[\{1-A_{jk}(i)\} \{1-R_{jk}(i)\}] (\mu_{ik}^{\mbox{\footnotesize MAR}} - \mu^b)
\nonumber\\
&=&(1-\pi_{ik}) \mu_{ik}^{\mbox{\footnotesize MAR}} + \pi_{ik} \mu^b, \nonumber \label{eq:estimand_r2b}
\end{eqnarray}
where $\pi_{ik} = \Pr\{R_{jk}(i)=0, A_{jk}(i)=0\}$ is the probability of all participants with missing values in Pattern B at time point $k$ under treatment $i$. 
Then, a consistent estimator for $\mu_{ik}^{\mbox{\footnotesize MAR-R2B}}$ can be given by:
\begin{equation} \label{eq:estimator_r2b}
\hat \mu_{ik}^{\mbox{\footnotesize MAR-R2B}} = (1- \hat \pi_{ik}) \hat \mu_{ik}^{\mbox{\footnotesize MAR}} + \hat \pi_{ik}  \hat \mu_{i}^b, 
\end{equation}
where $\hat \pi_{ik}$ is the observed proportion of participants with missing values and under Pattern B, for treatment group $i$ at time point $k$, $\hat \mu_{ik}^{\mbox{\footnotesize MAR}}$ is the estimator for the mean effect for treatment $i$ at time point $k$ under the MAR assumption (which may be estimated using a mixed model for repeated measures [MMRM] including all data up to the occurrence of intercurrent events), and $\hat \mu_{ik}^b$ is the sample mean of $Y_{j0}$ for patients randomized to treatment group $i$. \cite{zhang2020likelihood} suggested a variance estimator for $\hat \mu_{ik}^{\mbox{\footnotesize MAR-R2B}}$ when there are no missing values in Pattern A by assuming the independence of $\hat \pi_i$ and $\hat \mu_{i}^b$, which generally does not hold for the potential outcome in Equation (\ref{eq:RTB_PO}). A more general variance estimator is derived in Section \ref{sec:var}.

\subsection{J2R Imputation} \label{sec:J2R}
The intention of the J2R imputation is to impute the missing value in the experimental treatment group(s) using the potential outcome if the response returns immediately to that of the reference treatment arm. Similar to the condition for the R2B imputation, we assume that Assumption A3 holds. The potential outcome using the MAR imputation for missing values within Pattern A and the J2R imputation \citep{carpenter2013analysis,liu2017control,leurent2020reference} for missing values within Pattern B is given by
\begin{equation} \label{eq:J2R_PO}
Y_{jk}^{\mbox{\footnotesize MAR-J2R}}(i) = A_{jk}(i) R_{jk}(i) Y_{jk}(i) + A_{jk}(i) \{1-R_{jk}(i)\} Y_{jk}^{\mbox{\footnotesize MAR}}(i) + \{1-A_{jk}(i)\} \{1-R_{jk}(i)\} Y_{jk}^{\mbox{\footnotesize J2R}}(i),
\end{equation}
where $Y_{jk}^{\mbox{\footnotesize J2R}}(i) = Y_{jk}^{\mbox{\footnotesize MAR}}(i) - \mu_{ik}^{\mbox{\footnotesize MAR}} + \mu_{0k}^{\mbox{\footnotesize MAR}}$. 
Similar to the R2B imputation described in \cite{qu2022return}, a convenient J2R multiple imputation can be performed based on the imputed values under the MAR assumption:
\begin{equation} \label{eq:J2R_MI}
Y_{jk}^{\mbox{\footnotesize J2R,(r)}}(i) = Y_{jk}^{\mbox{\footnotesize MAR},(r)}(i)-\hat \mu_{ik}^{(r)}+\hat \mu_{0k}^{(r)}, \;\; \mbox{ for } R_{jk}(i)=0.
\end{equation}
Note the imputation given in Equation (\ref{eq:J2R_MI}) utilizes the within-subject variance-covariance matrix estimated from treatment group $i$, while \cite{carpenter2013analysis} used a hybrid variance-covariance matrix from treatment group $i$ and the reference group. There is no apparent reason for which variance-covariance matrix is better and the two variance-covariance matrices are generally similar, so the difference in the corresponding imputations should be small.

The corresponding estimand for the potential outcome defined in Equation (\ref{eq:J2R_PO}) is
\begin{eqnarray}
\mu_{ik}^{\mbox{\footnotesize MAR-J2R}} &=& E[A_{jk}(i) R_{jk}(i) Y_{jk}(i) + A_{jk}(i)\{1-R_{jk}(i)\}  Y_{jk}^{\mbox{\footnotesize MAR}}(i) \nonumber\\ && +  \{1-A_{jk}(i)\} \{1-R_{jk}(i)\}  \{Y_{jk}^{\mbox{\footnotesize MAR}}(i) - \mu_{ik}^{\mbox{\footnotesize MAR}} + \mu_{0k}^{\mbox{\footnotesize MAR}}\}] 
\nonumber \\
&=& (1-\pi_{ik}) \mu_{ik}^{\mbox{\footnotesize MAR}} +  \pi_{ik} \mu_{0k}^{\mbox{\footnotesize MAR}}. \nonumber \label{estimand_J2R}
\end{eqnarray}
Then, a straightforward estimator for $\mu_{ik}^{\mbox{\footnotesize MAR-J2R}}$ can be constructed as
\begin{equation} \label{eq:estimator_j2r}
\hat \mu_{ik}^{\mbox{\footnotesize MAR-J2R}} = (1 - \hat \pi_{ik}) \hat \mu_{ik}^{\mbox{\footnotesize MAR}} +  \hat \pi_{ik} \hat \mu_{0k}^{\mbox{\footnotesize MAR}}. 
\end{equation}
There are two attractive features of the J2R estimator: 
\begin{enumerate}
    \item  The equation can also apply to the reference group in which the estimator in Equation \eqref{eq:estimator_j2r} becomes the MAR estimator.
    \item When there is no treatment effect, the estimands for the treatment and placebo groups are the same.  
\end{enumerate}

\cite{mehrotra2017missing} provided a ``control-based mean imputation" estimator without actually performing the imputation:
\[
\hat \mu_{ik}^{M} = (1-\hat \pi_{ik})  \hat \mu_{ik}^c +  \hat \pi_{ik} \hat \mu_{0k}^{\mbox{\footnotesize MAR}},
\]
where $\hat \mu_{ik}^c$ is an estimator for $E[Y_{jk}(i)|R_{jk}(i)=1]$, the mean for the completers (assuming no participants have missing values in Pattern A) for treatment group $i$. However, $\hat \mu_{ik}^{M}$ is generally not a consistent estimator for $\mu_{ik}^{\mbox{\footnotesize J2R}}$ as $\hat \mu_{ik}^{c}$ is not a consistent estimator for $\mu_{ik}^{\mbox{\footnotesize MAR}}$  [unless  $R_{jk}(i) \ind Y_{jk}(i)$, i.e., \emph{missing completely at random} (MCAR)]. If there is no treatment effect for $i=1,2,\ldots, I$, under the MAR but not MCAR assumption, $E\left(\hat \mu_{ik}^{M}\right) \ne E\left(\hat \mu_{0k}^{}\right)$ while $E\left(\hat \mu_{ik}^{\mbox{\footnotesize MAR-J2R}}\right) = E\left(\hat \mu_{0k}\right)$. Therefore, $\hat \mu_i^{\mbox{\footnotesize MAR-J2R}}$ is a more robust estimator of $\mu_{ik}^{\mbox{\footnotesize MAR-J2R}}$ than $\hat \mu_{ik}^{M}$.

\subsection{PW Imputation} \label{sec:PW}
\cite{wang2023statistical} proposed a PW method for placebo-controlled studies which assumes that the treatment effect will be ``washed out" from the time data are missing. For this approach, the missing values for the placebo group are imputed under the MAR assumption, while the missing values for the experimental treatment group(s) are imputed using data at baseline and final time point of interest (no intermediate outcomes are included) from participants in placebo group. In this case, this method essentially uses only placebo ``completers" to impute the missing values in the experimental group(s), which does not reflect the placebo response for the whole (randomized) population. Therefore, we make a small adjustment to the PW imputation by using all placebo data but only including the baseline data for experimental group(s) in the imputation of the missing values for the experimental group(s). 

It is assumed that $E\{Y_{jk}(0,0)|\bd X_j\} = \beta_{0k,0}+\bd X_{j}' \bd \beta_{0k,1}$. The parameter estimator $\hat \beta_{0k,0}$ and $\hat{\bd \beta}_{0k,1}$ can be obtained from an MMRM that includes the interaction of baseline variable $X_{j}$ and the discrete time point $t_k$ using data in the placebo treatment group $i=0$. The potential outcome using MAR imputation for missing values within Pattern A and using the PW imputation for missing values within Pattern B is given by
\begin{eqnarray} 
Y_{jk}^{\mbox{\footnotesize MAR-PW}}(i) &=& A_{jk}(i) R_{jk}(i) Y_{jk}(i) + A_{jk}(i) \{1-R_{jk}(i)\} Y_{jk}^{\mbox{\footnotesize MAR}, A=1}(i) \nonumber\\ && + \{1-A_{jk}(i)\} \{1-R_{jk}(i)\} Y_{jk}^{\mbox{\footnotesize PW}}(i), \;\;\;\; i \ne 0,
\label{eq:J2R_PW} \nonumber
\end{eqnarray}
where $Y_{jk}^{\mbox{\footnotesize MAR}, A=1}(i)$ is the potential outcome for data imputed under the MAR assumption based on participants with $A_{jk}(i)=1$ and $Y_{jk}^{\mbox{\footnotesize PW}}(i) = \hat \beta_{i0,0} + \bd X_j' \hat{\bd \beta}_{i0,1} + \epsilon_{jk}$ is the potential outcome with the PW imputation. 
With Assumption A3, we have 
\begin{eqnarray} 
E\{Y_{jk}^{\mbox{\footnotesize MAR-PW}}(i)\} &=& E[A_{jk}(i) R_{jk}(i) Y_{jk}(i) + A_{jk}(i) \{1-R_{jk}(i)\} Y_{jk}^{\mbox{\footnotesize MAR}, A=1}(i)] \nonumber \\ && + E[\{1-A_{jk}(i)\} \{1-R_{jk}(i)\} Y_{jk}^{\mbox{\footnotesize PW}}(i)]
\nonumber \\ &=&
E[R_{jk}(i) Y_{jk}(i) + \{1-R_{jk}(i)\} Y_{jk}^{\mbox{\footnotesize MAR}, A=1}(i)|A_{jk}(i) =1] \Pr\{A_{jk}(i) =1\} \nonumber \\ && + E[
\{Y_{jk}^{\mbox{\footnotesize PW}}(i)|A_{jk}(i)=0, R_{jk}(i)=0\} \Pr\{A_{jk}(i)=0, R_{jk}(i)=0\}
\nonumber \\&=&
 (1-\pi_{ik})\mu_{ik}^{\mbox{\footnotesize MAR}, A=1} +  \pi_{ik} \{\beta_{0k,0} + ({\bd \nu}_{ik}^{A=0})' {\bd \beta}_{0k,1} \}, \label{eq:estimand_pw} \nonumber
\end{eqnarray}
where $\mu_{ik}^{\mbox{\footnotesize MAR}, A=1} = E[R_{jk}(i) Y_{jk}(i) + \{1-R_{jk}(i)\} Y_{jk}^{\mbox{\footnotesize MAR}, A=1}(i)|A_{jk}(i) =1]$ and ${\bd \nu}_{ik}^{A=0}=E\{\bd X_j|A_{jk}(i)=0\}$ is the expected value of baseline covariates for participants belonging to Pattern B at time point $k$ in treatment group $i$.

Then, a straightforward estimator for $\mu_{ik}^{\mbox{\footnotesize MAR-PW}}$ can be constructed as
\begin{equation} \label{eq:estimator_pw} 
\hat \mu_{ik}^{\mbox{\footnotesize MAR-PW}} = (1 - \hat \pi_{ik}) \hat \mu_{ik}^{\mbox{\footnotesize MAR}, A=1} +  \hat \pi_{ik} \{\hat \beta_{i0,0} + (\hat{\bd \nu}_{ik}^{A=0})' \hat {\bd \beta}_{i0,1} \},
\end{equation}
where $\hat {\bd \nu}_{ik}^{A=0}$ is the mean of baseline covariates for participants with $A_{jk}(i)=0$. 

Note that the PW estimator does not have the same attractive properties as those of the J2R estimator. If the experimental treatments have no effect, the estimand for treatment group $i$ ($i \ne 0$) is not equal to the estimand for the placebo group under MAR but not MCAR. 

\subsection{RD Imputation} \label{sec:rd}
When there is a sufficient number of RDs, the RD imputation is commonly used to impute the potential outcome with intercurrent events handled by the treatment policy strategy. Again, missing values are imputed according to the pattern of missingness. If $A_{jk}(i)=1$ and $Y_{jk}(i)$ is missing, the observed data up to time point $k$ (i.e., $R_{jl}(i)=1, l=1,2,\ldots,k$) for those participants with $A_{jk}(i) = 1$ are used to impute $Y_{jk}(i)$ under the MAR assumption. The missing values of $Y_{jk}(i)$ for participants with $A_{jk}(i)=0$ can be imputed using the retrieved dropouts (data on subjects with $A_{jk}(i)=0$ and $R_{jk}(i)=1$), which is often called the RD imputation. 

The potential outcome under the MAR imputation for missing values within Pattern A and following the distribution of the RDs for missing values within Pattern B is given by
\begin{eqnarray}
Y_{jk}^{\mbox{\footnotesize MAR-RD}}(i) &=& A_{jk}(i)[R_{jk}(i) Y_{jk}(i) + \{1-R_{jk}(i)\} Y_{jk}^{\mbox{\footnotesize MAR}, A=1}(i)] \nonumber\\ && +(1-A_{jk}(i)) [ R_{jk}(i) Y_{jk}(i) + \{1-R_{jk}(i)\}  Y_{jk}^{\mbox{\footnotesize CovMAR}, A=0}(i)], 
\label{eq:RD_PO}
\end{eqnarray}
where $Y_{jk}^{\mbox{\footnotesize MAR}, A=1}(i)$ is the potential outcome under the MAR assumption based on subjects with $A_{jk}(i)=1$, and $Y_{jk}^{\mbox{\footnotesize CovMAR}, A=0}(i)$ is the potential outcome for data imputed assuming that the probability of missing values depends only on the baseline covariates for participants with $A_{jk}(i)=0$. 

Let the conditional distribution of $Y_{jk}(i)$ given baseline covariates be:
\begin{equation} \label{eq:rd_A1} \nonumber
E\{Y_{jk}(i)|A_{jk}(i)=0, \bd X_j\} = \beta_{ik,0}^- + {\bd X}_j' \bd \beta_{ik,1}^-.
\end{equation}
The estimand corresponding to the potential outcome in Equation (\ref{eq:RD_PO}) is
\begin{eqnarray}
\mu_{ik}^{\mbox{\footnotesize MAR-RD}} 
&=& E\left(A_{jk}(i)\left[R_{jk}(i) Y_{jk}(i) + \{1-R_{jk}(i)\} Y_{jk}^{\mbox{\footnotesize MAR}, A=1}(i)\} \right] \right. \nonumber\\ && \left. +\{1-A_{jk}(i)\} \left[  R_{jk}(i) Y_{jk}(i) + \{1-R_{jk}(i)\} Y_{jk}^{\mbox{\footnotesize CovMAR}, A=0}(i) \right] \right) 
\nonumber \\ &=&
E[R_{jk}(i) Y_{jk}(i) + \{1-R_{jk}(i)\} Y_{jk}^{\mbox{\footnotesize MAR}, A=1}(i)|A_{jk}(i)=1] \Pr\{A_{jk}(i)=1\} \nonumber\\ && + E[ R_{jk}(i) Y_{jk}(i) +  \{1-R_{jk}(i)\}Y_{jk}^{\mbox{\footnotesize CovMAR}, A=0}(i)]|A_{jk}(i)=0] \Pr\{A_{jk}(i)=0\} \nonumber\\
&=& \phi_{ik} \mu_{ik}^{\mbox{\footnotesize MAR}, A=1}  + (1-\phi_{ik}) (\beta_{ik,0}^- + ({\bd \nu}_{ik}^{A=0})' \bd \beta_{ik,1}^-)  \nonumber,
\end{eqnarray}
where $\phi_{ik} =\Pr\{A_{jk}(i)=1\}$, and $\mu_{ik}^{\mbox{\footnotesize MAR}, A=1}$ and ${\bd \nu}_{ik}^{A=0}$ are define as in Section \ref{sec:PW}.
Then, a consistent estimator for $\mu_{ik}^{\mbox{\footnotesize MAR-RD}}$ is given by
\begin{equation} \label{eq:estimator_rd}
\hat \mu_{ik}^{\mbox{\footnotesize MAR-RD}} = \hat \phi_{ik} \hat \mu_{ik}^{\mbox{\footnotesize MAR}, A=1}  + (1-\hat\phi_{ik}) \{\hat \beta_{ik,0}^- + (\hat {\bd \nu}_{ik}^{A=0})' \hat {\bd \beta}_{ik,1}^- \},
\end{equation}
where $\hat \phi_{ik}$ is the observed proportion of adherers in treatment group $i$, $\hat \mu_{ik}^{\mbox{\footnotesize MAR}, A=1}$ is the estimator for $\mu_{ik}^{\mbox{\footnotesize MAR}, A=1}$ (which can be obtained by fitting an MMRM for non-missing data up to time point $k$ for participants with $A_{jk}(i)=1$), $\hat {\bd \nu}_{ik}^{A=0}$ is the mean of baseline covariates for participants with $A_{jk}(i)=0$, and $\hat \beta_{ik,0}^-$ and $\hat {\bd \beta}_{ik,1}^-$ are the intercept and slope, respectively, from the linear regression of $Y$ on $\bd X$ for participants with $A_{jk}(i)=0$ and $R_{jk}(i)=1$.

If there are no missing values for participants with $A_{jk}(i)=1$, all missing values are imputed using the RD imputation. Then, we have
\begin{eqnarray}
\mu_{ik}^{\mbox{\footnotesize RD}}
&=& E\{R_{jk}(i)Y_{jk}(i) + (1-R_{jk}) Y_{jk}^{\mbox{\footnotesize CovMAR}, A=0}(i)\} \nonumber\\
&=& E\{Y_{jk}(i)|R_{jk}(i)=1\}\Pr\{R_{jk}(i)=1\} + E\{Y_{jk}^{\mbox{\footnotesize CovMAR}, A=0}(i)|R_{jk}(i)=0\}\Pr\{R_{jk}(i)=0\} \nonumber\\
&=& (1-\pi_{ik}) \mu_{ik}^{R=1} + \pi_{ik} \{\beta_{ik,0}^- + (\bd \nu_{i}^{R=0})' \bd \beta_{ik,1}^-\} \nonumber,
\end{eqnarray}
where $\mu_{ik}^{R=1}=E\{Y_j(i)|R_{jk}(i)=1\}$ and $\bd \nu_{ik}^{R=0} = E\{\bd X_j|R_{jk}(i)=0\}$. Then, an estimator for $\mu_{ik}^{\mbox{\footnotesize RD}}$ is
\[
\hat \mu_i^{\mbox{\footnotesize RD}} = (1 - \hat \pi_{ik}) \hat \mu_{ik}^{R=1} + \hat\pi_{ik} \{\hat \beta_{ik,0}^{-} + (\hat {\bd \nu}_{ik}^{R=0})' \hat \beta_{ik,1}^{-}\},
\]
where $\hat \mu_{ik}^{R=1}$ is the sample mean of $Y_j$ among participants without missing values at time point $k$ for treatment group $i$ and $\hat {\bd \nu}_{ik}^{R=0}$ is the mean of $\bd X_j$ for participants with missing values at time point $k$ in treatment group $i$. 

\subsection{The $\delta$-Adjusted and Tipping Point Sensitivity Analyses} \label{sec:delta_adjusted}
For the estimator based on the MAR assumption and the estimators based on the pattern-mixture models described in Sections \ref{sec:r2b} through \ref{sec:rd}, sensitivity analyses can be performed by adding additional penalty parameters on imputed missing values. This method is often called the $\delta$-adjusted method. The tipping point is the point at which the sensitivity parameter makes the treatment effect become non-significant. These penalty values can be added to the experimental treatment group alone (one-way $\delta$-adjusted and tipping point analyses) or to both the reference and experimental treatment groups (two-way $\delta$-adjusted and tipping point analyses). This section discusses the two-way $\delta$-adjusted analysis, as the one-way $\delta$-adjusted analysis is a special case. 

We drop the subscript $k$ to simplify the notation. Let $\hat \mu_i^*$ be the estimator for treatment group $i$ based on a certain imputation method and $\hat \tau_i$ be the observed proportion of participants with missing values that are imputed under the MAR assumption (i.e., $\hat \tau_i$ is an estimator for $\Pr\{R_{jk}(i)=0, A_{jk}(i)=1$\}). The $\delta$-adjusted estimator is
\[
\hat \mu_i^{*\delta} = \hat \mu_i^* + \delta_i (\hat \pi_i+ \hat\tau_i), \quad i=0,1, \ldots, I.
\]
The treatment difference is
\begin{equation} \label{eq:est_delta}
\hat \mu_i^{*\delta} - \hat \mu_0^{*\delta} = (\hat \mu_i^* - \hat \mu_0^*) + \delta_i  (\hat \pi_i + \hat\tau_i) - \delta_0 (\hat \pi_0 + \hat \tau_0), \quad i=1,2,\ldots, I.
\end{equation}
The variance estimator $\widehat{Var}\left(\hat \mu_i^{*\delta} - \hat \mu_0^{*\delta} \right)$, which can be easily obtained using the method described in Section \ref{sec:var}, is a quadratic function of $\delta_0$ and $\delta_1$, expressed as 
\begin{equation} \label{eq:est_tp_var}
\widehat{Var}\left(\hat \mu_i^{TP} - \hat \mu_0^{TP} \right) = a_0 \delta_0^2 + a_i \delta_i^2 + b_{i0} \delta_0 + b_{i1} \delta_i + c_i.
\end{equation}
The tipping points (boundary) for the significance boundary at the two-sided $\alpha$ level are determined by
\begin{equation} \label{eq:tp_boundary}
    (\hat \mu_i^{*\delta} - \hat \mu_0^{*\delta}) + z_{1-\alpha/2} \sqrt{\widehat{Var}\left(\hat \mu_i^{*\delta} - \hat \mu_0^{*\delta} \right)} = 0.
\end{equation}
Substituting Equations (\ref{eq:est_delta}) and (\ref{eq:est_tp_var}) into Equation  (\ref{eq:tp_boundary}), we have
\[
z_{1-\alpha/2}^2 (a_0 \delta_0^2 + a_i \delta_i^2 +  b_{0} \delta_0 +  b_i \delta_i + c_i) =  [(\hat \mu_i - \hat \mu_0) + \{\delta_i (\hat \pi_i + \hat \tau_i) - \delta_0 (\hat \pi_0 + \hat \tau_0)\}]^2.
\]
Then, the tipping points regarding $\delta_0$ and $\delta_i$ can be obtained by solving the above quadratic equation, which is generally a segment of an ellipse.  

\subsection{Adjusting for baseline covariates} \label{sec:base_adj}
Some important baseline covariates are often adjusted for in the analysis model. It is easy to adjust for the baseline covariates in the linear model with the ``complete" data after imputation, but the adjustment for baseline covariates in estimators described in Sections \ref{sec:r2b} through \ref{sec:rd} requires a different approach. We use the property of the Gaussian conditional distribution to construct baseline adjusted estimators, an approach similar to that described in \cite{jiang2019robust}. 

In this section, we drop the subscript $k$ to simplify the notation. Let $\hat {\bd \mu} = (\hat \mu_0, \hat \mu_1, \ldots, \hat \mu_I)'$ denote a vector of consistent estimators for a desired estimand of $\tilde{\bd \mu} = (\tilde \mu_{0}, \tilde \mu_{1}, \ldots, \tilde \mu_{I})$, which is the mean for the response variable under a certain method of handling intercurrent events and missing values. Note that $\hat{\bd \mu}$ can be the estimator based on the R2B, J2R, PW, or RD pattern-mixture model.  Let $\hat {\bd \nu} = (\hat {\bd \nu}_{0}', \hat {\bd \nu}_{1}', \ldots, \hat {\bd \nu}_{I}')'$ be the vector of the means for the baseline covariates $\bd X_j$ for treatment group $i$ ($i=0,1,\ldots,I$). For a sufficiently large sample size, we can assume $\hat {\bd \mu}$ and $\hat{\bd \nu}$ approximately follow a normal distribution:
\begin{equation} \nonumber
    \left[\begin{array}{c} \hat {\bd \mu} \\ \hat {\bd \nu} \end{array} \right] \approx N\left( \left[\begin{array}{c} \tilde{\bd \mu} \\ \bd \nu \end{array} \right], \left[\begin{array}{cc} \Sigma_{yy} & \Sigma_{yx} \\ \Sigma_{yx}' & \Sigma_{xx}\end{array} \right]
    \right).
\end{equation}
Let $\bar {\bd X}^L = (\underbrace{\bar{\bd X}', \bar{\bd X}', \ldots, \bar{\bd X}'}_{\mbox{repeat } I+1 \mbox{ times}})'$, 
where $\bd{\bar X} = n^{-1}\sum_{j=1}^n \bd X_j$ denotes the mean of the baseline covariate for all participants across treatment groups. Since the conditional distribution of $\hat {\bd \mu} |\hat {\bd \nu} = \bar {\bd X}^L$ is
\[
\hat {\bd \mu}|_{\hat {\bd \nu} = \bar {\bd X}^L} \approx  N\left(\tilde{\bd \mu} + \Sigma_{yx}\Sigma_{xx}^{-1}(\bar{\bd X}^L - \bd \nu), \Sigma_{yy} - \Sigma_{yx}\Sigma_{xx}^{-1}\Sigma_{yx}'\right),
\]
then the adjusted estimator for $\bd \mu_y$ is given by
\begin{equation} \label{eq:mu_hat_adj} \nonumber
\hat {\bd \mu}_{adj} = \hat {\bd \mu} + \hat \Sigma_{yx} \hat\Sigma_{xx}^{-1}(\bar{\bd X}^L - \hat{\bd \nu}),
\end{equation}
and the corresponding variance estimator (conditional on $\bd X$) for $\hat {\bd \mu}_{adj}$ is
\begin{equation} \label{eq:mu_hat_adj_var_cond} \nonumber
\hat V(\hat {\bd \mu}_{adj}|\bd X)= \hat \Sigma_{yy} - \hat \Sigma_{yx} \hat\Sigma_{xx}^{-1} \hat \Sigma_{yx}',
\end{equation}
where $\hat \Sigma_{yy}$, $\hat \Sigma_{yx}$, and $\hat \Sigma_{xx}$ are the estimators for $\Sigma_{yy}$, $\Sigma_{yx}$, and $\Sigma_{xx}$, respectively. 

Note
\begin{eqnarray} 
E(\hat {\bd \mu}_{adj}|\bd X)
&=& E(\hat {\bd \mu}|\bd X) + 
\Sigma_{yx} \Sigma_{xx}^{-1} (\bar{\bd X}^L - \hat{\bd \nu}) \nonumber\\
&=& \tilde {\bd \mu} + \Sigma_{yx} \Sigma_{xx}^{-1} (\hat{\bd \nu} - \bd \nu) + 
\Sigma_{yx} \Sigma_{xx}^{-1} (\bar{\bd X}^L - \hat{\bd \nu}) \nonumber\\
&=& \tilde {\bd \mu} - \Sigma_{yx} \Sigma_{xx}^{-1} \bd \nu + 
\Sigma_{yx} \Sigma_{xx}^{-1} \bar{\bd X}^L \nonumber\\
&=& \tilde {\bd \mu} - \Sigma_{yx} \Sigma_{xx}^{-1} \bd \nu + 
\Sigma_{yx} \Sigma_{xx}^{-1} A' \bar{\bd X}, \nonumber
\end{eqnarray}
where $A = [I_{m \times m}, I_{m \times m}, \cdots, I_{m \times m}],$ and $I_{m \times m}$ is the $m\times m$ identity matrix.
Then, the \emph{unconditional} variance estimator is 
\begin{eqnarray} \nonumber
\hat V(\hat {\bd \mu}_{adj}) &=& \hat V(\hat {\bd \mu}_{y,adj}|\bd X) + \hat V\{E(\hat {\bd \mu}_{y,adj}|\bd X)\} \nonumber \\ &=& \hat \Sigma_{yy} - \hat \Sigma_{yx} \hat\Sigma_{xx}^{-1} \hat \Sigma_{yx}' + n^{-1}  \hat \Sigma_{yx} \hat\Sigma_{xx}^{-1}A' S_x^2 A \hat\Sigma_{xx}^{-1} \hat \Sigma_{yx}', \nonumber
\end{eqnarray}
where $\bd S_x^2$ is the sample variance of $\bd X$. 

The baseline adjustment in this section allows between-treatment heterogeneity for the relationship between baseline $\bd X_j$ and the longitudinal outcome $\bd Y_j$. Therefore it achieves the efficiency similar to that using the analysis of heterogeneous covariance model \citep{ye2022toward} or the improved MMRM \citep{wang2023improving} for the ``complete" data after imputation, if all randomization strata are included in the baseline covariates.  

\subsection{Variance estimation for parameters used for constructing R2B, J2R, RD, and PW estimators} \label{sec:var}
In Sections \ref{sec:r2b} through \ref{sec:rd}, we described the point estimator $\hat {\bd \mu}_y$ for the mean response of each treatment group. In Section \ref{sec:delta_adjusted}, we discussed the $\delta$-adjusted sensitivity analysis. In this section, we describe the method to estimate the variance for these estimators. Note that $\hat {\bd \mu}_y$ is a function ($h$) of $\hat {\bd \theta}$, the estimators for the relevant parameters $\bd \theta$, and can be expressed as
\begin{equation} \label{eq:h}
\hat {\bd \mu} = h\bd (\hat{\bd \theta}, \bd \delta), 
\end{equation}
where $\bd \delta = (\delta_0, \delta_1, \ldots, \delta_I)'$ is a set of sensitivity parameters introduced in Section \ref{sec:delta_adjusted}. The estimator $\hat {\bd \mu}$ without a sensitivity parameter is a special case of Equation (\ref{eq:h}) by setting $\bd \delta = \bd 0$. Generally, $\bd \theta$ is estimated from an unbiased estimating equations (UEE) system:
\begin{equation} \label{eq:psi} \nonumber
\sum_{j=1}^n \bd \psi(\bd X_j, \bd Y_j^o, R_j, T_j; \bd \theta) = \bd 0,
\end{equation}
where the superscript $o$ for $\bd Y$ indicates the non-missing portion of the data. Here we assume there is no missing value at baseline.

Then, the sandwich variance estimator for $\hat{\bd \theta}$ is
\begin{equation}\label{eq:sandwich} \nonumber
\hat V(\hat{\bd \theta}) = \left( \sum_{j=1}^n \frac{\partial {\hat {\bd \psi}_j}}{\partial \bd \theta'} \right)^{-1} \left( \sum_{j=1}^n \hat {\bd \psi}_j \hat {\bd \psi}'_j \right)
\left\{ \left(\sum_{j=1}^n \frac{\partial {\hat {\bd \psi}_j}}{\partial \bd \theta'} \right)^{-1} \right\}',
\end{equation}
where
$\hat {\bd \psi}_j = \bd \psi(\bd X_j, \bd Y_j^o, R_j, T_j; \hat{\bd \theta})$. It follows that
\begin{equation}\label{eq:delta_method}
\hat V(\hat{\bd \mu}) = \left[\frac{\partial \bd h (\hat{\bd \theta}, \bd\delta)}{\partial \bd \theta} \right] \hat V(\hat{\bd \theta}) \left[\frac{\partial \bd h (\hat{\bd \theta}, \bd\delta)}{\partial \bd \theta} \right]'.
\end{equation}
We suggest $\bd \nu$ be part of $\bd \theta$, so the variance for $\hat{\bd \nu}$ and the covariance between $\hat {\bd \nu}$ and $\hat {\bd \mu}$ can be estimated similarly using the delta-method. Then, the variance for $\hat{\bd \mu}_{adj}$, the estimator adjusting for baseline covariates described in Section \ref{sec:base_adj}, can be easily obtained. Next, we will describe how to construct the UEEs for R2B, J2R, PW, and RD pattern mixture models. 


\subsubsection{Variance Estimation for R2B and J2R Pattern-Mixture Models} \label{sec:var_r2b_j2r}
Let $\bd Y_j(i)$, the potential outcome for a vector of the longitudinal response for participant $j$ randomized to treatment $i$, follows a multivariate normal distribution
\[
\bd Y_j(i) \sim N\left(\bd X_j' \bd \beta_i, \Sigma_i\right),
\]
where $\bd X_j$ is a vector for the fixed effects, $\bd \beta_i = \left(\beta_{i1,0}, \bd \beta_{i1,1}', \beta_{i2,0}, \bd \beta_{i2,1}',\ldots, \beta_{iK,0}, \bd \beta_{iK,1}'\right)'$ is the regression coefficient of the longitudinal outcome $Y_j$ under treatment $i$, and $\Sigma_i$ is the within-subject variance-covariance matrix for treatment $i$. With known variance-covariance, the maximum likelihood estimator for 
$\bd \beta_i$ can be obtained through the UEE function:
\[
\bd \psi_{\beta i}(\bd X_j, \bd Y_j, T_j; \bd \beta_i) = I(T_j=i) \cdot \{(\mathbb{X}_j' \Sigma_i^{-1} \mathbb X_j) \bd\beta_i - (\mathbb{X}_j' \Sigma_i^{-1} \bd Y_j)\},
\]
where
\[
\mathbb X_j = \left[ \begin{array}{cccccccc} 1 & \bd X_j' & & & & & & \\ & & 1 &  \bd X_j' & & & & \\ & & & & \ldots & \ldots & & \\
& & & & & & 1 & \bd X_j' 
\end{array} \right]_{K \times \{K(m+1)\}}.
\]
If there are missing values, we assume ignorable missingness for the potential outcome $Y_j(i)$:
$$Y_j(i) \perp R_j(i) | \bd X_j, Y_j^o(i).$$ 
Then, the above UEE function can be modified as
\begin{equation} \label{eq:psi_mar}
\bd \psi_{\beta i}(X_j, \bd Y_j^o, T_j; \bd \beta_i) = I(T_j=i) \cdot \{(\mathbb X_j^o)' (\Sigma_i^o)^{-1} \mathbb X_j^o \bd\beta_i - (\mathbb X_j^o)' (\Sigma_i^o)^{-1} \bd Y_j^o\},
\end{equation}
where the superscript ``$o$" for $\mathbb X_j$ and $\Sigma_i$ indicates the portion of $\mathbb X_j$ and $\Sigma_i$ corresponding to non-missing $Y_j$ values, respectively. 
The proportion of participants with missing values that are imputed using a special missing not at random pattern for treatment group $i$, $\bd \pi_i = (\pi_{i1}, \pi_{i2},\ldots, \pi_{iK})'$, can be estimated through the UEE function of
\begin{equation} \label{eq:psi_2} \nonumber
\psi_{\pi i}(\bd R_j, \bd A_j, T_j; \bd \pi) = I(T_j=i) \cdot \{(1-\bd A_j)(1-\bd R_j) - \bd \pi_{i}\},
\end{equation}
where $\bd R_j = (R_{j1}, R_{j2}, \ldots, R_{jK})'$. 
Similarly, the proportion of participants with missing values that are imputed under the MAR assumption for treatment group $i$, $\bd \tau_i = (\tau_{i1}, \tau_{i2},\ldots, \tau_{iK})'$, can be estimated through the UEE function of
\begin{equation} \label{eq:psi_tau} \nonumber
\bd \psi_{\tau i}(\bd A_j, \bd R_j, T_j; \bd \pi) = I(T_j=i) \cdot \{\bd A_j(1-\bd R_j) - \bd \tau_i\}.
\end{equation}
The mean baseline value for treatment group $i$ can be estimated by the UEE function of
$$
\bd \psi_{\nu i}(\bd X_j, T_j; \bd \nu_{i}) = I(T_j=i) \cdot (\bd X_j - \bd \nu_{i}).
$$
Then, the UEE function for estimating $\bd \theta_i = (\bd \beta_i', \bd \pi_i, \bd \tau_i, \bd \nu_{i}')'$ is given by
$$ 
\bd \psi_i(\bd X_j, \bd Y_j^o, \bd R_j, \bd Q_j, T_j; \bd \theta_i) = \left[ \begin{array}{c} \bd \psi_{\beta i}(\bd X_j, \bd Y_j^o, T_j; \bd \beta_i) \\ \bd \psi_{\pi i}(\bd R_j, T_j; \bd \pi_i) \\
\bd \psi_{\tau i}(\bd Q_j, T_j; \bd \tau_i) \\
\bd \psi_{\nu i}(\bd X_j, T_j; \bd \nu_{i})
\end{array} \right].
$$
The variance-covariance matrix for $\hat {\bd \theta}_i$ can be estimated using the sandwich variance estimation method. The variance for the R2B and J2R estimators can be obtained using the delta-method in Equation (\ref{eq:delta_method}) with the corresponding $h$ functions.

Note that $\{\hat {\bd \theta}_i: i=0,1,\ldots,I\}$ are independent across treatment arms. Additionally, $\hat \beta_i$  {can be obtained from procedures or functions found in} commonly used statistical software packages {such as SAS\textsuperscript{\textregistered} (PROC MIXED) software or R (lme() function in `nlme' package, geeglm() function in `geepack' package) \citep{r2022}}. The values of $\bd \psi_{\beta i}(X_j, \bd Y_j^o, T_j; \hat{\bd \beta}_i)$ and $\frac{\partial \bd \psi_{\beta i}(X_j, \bd Y_j^o, T_j; \hat{\bd \beta}_i)}{\partial \bd \beta}$ can often be obtained as optional outputs. Therefore, the sandwich variance estimation can be obtained without customized complex numeric computation. 

\subsubsection{Variance Estimation for the RD Pattern-Mixture Model} \label{sec:var_rd}
In this section, we consider the estimator $\hat \mu_{iK}^{\mbox{\footnotesize MAR-RD}}$, the estimator at the final time point $K$ based on the RD imputation, which is generally needed. Variance estimation for the RD estimator at other time points can be similarly obtained. Let $\bd \beta_i^+ = \left[\beta_{i1,0}^+, (\bd \beta_{i1,1}^+)', \beta_{i2,0}^+, (\bd \beta_{i2,1}^+)',\ldots, \beta_{iK,0}^+, (\bd \beta_{iK,1}^+)'\right]'$ denote the regression coefficients for the longitudinal outcome $\bd Y_j$ in participants in Pattern A. Therefore, under the MAR assumption, $\bd \beta_i^+$ can be estimated through the maximum likelihood with the following UEE function:
\begin{equation}\label{eq:psi_betaplus_mar}
\bd \psi_{\beta+, i}(\bd X_j, \bd Y_j^o, A_j, R_j, T_j; \bd \beta_{i}^+) = I(T_j=i, A_j=1) \cdot \{(\mathbb X_j^o)' (\Sigma_i^o)^{-1} \mathbb X_j^o \bd\beta_i^+ - (\mathbb X_j^o)' (\Sigma_i^o)^{-1} \bd Y_j^o\}.
\end{equation}
The coefficients for regressing $Y_j$ on $\bd X_j$ for participants in Pattern B, $\bd \beta_{iK}^- = \left[\beta_{iK,0}^-, \left(\bd \beta_{iK,1}^-\right)'\right]'$, can be estimated through the UEE function of 
\[
\bd \psi_{\beta-, iK}(\bd X_j, \bd Y_j^o, A_j, R_{jK}, T_j; \bd \beta_i^-) = I(T_j=i, A_j=0, R_{jK}=1) \cdot \left\{(1, \bd X_j')' (1, \bd X_j') \bd \beta_{iK}^- - (1, \bd X_j')' Y_{jK} \right\}.
\]
The proportion of participants in Pattern A, $\phi_{iK}$, can be estimated through the UEE function of
\[
\bd \psi_{\phi iK}(A_{jK}, T_j; \phi_{iK}) = I(T_j=i) \cdot (\phi_{iK} - A_{jK}).
\]
The baseline mean value for participants in Pattern B (i.e., with $A_{jK}(i)=0$) can be estimated through the UEE function of
\[
\bd \psi_{\nu iK, A=0}(\bd X_j, A_{jK}, R_{jK}, T_j; \bd \nu_{iK}^{A=0}) = I(T_j=i, A_{jK}=0, R_{jK}=0) \cdot (\bd \nu_{iK}^{A=0} - \bd X_j).
\]

Then, the parameter $\bd \theta_i = \left[ \left(\bd \beta_{i}^+\right)', \left(\bd \beta_{i}^-\right)', \phi_{iK}, \tau_{iK}, \left(\nu_{ik}^{A=0}\right)' \right]'$ for the RD pattern mixture model can be estimated by the following UEE functions:
$$ 
\bd \psi_i(\bd X_j, \bd Y_j^o, A_j, R_j, T_j; \bd \theta_i) = \left[ \begin{array}{c} \bd \psi_{\beta+, i}(X_j, \bd Y_j^o, A_j, T_j; \bd \beta_i^+) \\ \bd \psi_{\beta-, iK}(X_j, \bd Y_j^o, A_{jK}, T_j; \bd \beta_i^-) \\ \psi_{\phi iK}(A_{jK}, R_{jK}, T_j; \phi_{iK}) \\ \psi_{\tau iK}(Q_{jK}, T_j; \bd \tau_i) \\
\bd \psi_{\nu iK, A=0}(\bd X_j, A_{jK}, \bd R_{jK}, T_j; \bd \nu_{iK}^{A=0}) \\ \bd \psi_{\nu i}(\bd X_j, T_j; \bd \nu_{i})
\end{array} \right],
$$
where $\psi_{\tau iK}$ is the $K^{\mbox{\footnotesize th}}$ component of $\bd \psi_{\tau i}$. 

\subsubsection{Variance Estimation for the PW Pattern-Mixture Model}
Note that the parameters used for the PW estimands are a subset of the parameters in the J2R and RD estimands. Therefore, without repeating the derivations in Sections \ref{sec:var_r2b_j2r} and \ref{sec:var_rd}, the UEEs used for the estimation of the PW pattern-mixture model are given by
$$ 
\bd \psi_i(\bd X_j, \bd Y_j^o, \bd R_j, \bd Q_j, T_j; \bd \theta_i) = \left[ \begin{array}{c} \bd \psi_{\beta i}(\bd X_j, \bd Y_j^o, T_j; \bd \beta_i) \\ \bd \psi_{\pi i}(\bd R_j, T_j; \bd \pi_i) \\
\bd \psi_{\nu i}(\bd X_j, T_j; \bd \nu_{i})  \\
\bd \psi_{\beta+, i}(X_j, \bd Y_j^o, A_j, T_j; \bd \beta_i^+) \\
\bd \psi_{\nu iK, A=0}(\bd X_j, A_{jK}, \bd R_{jK}, T_j; \bd \nu_{iK}^{A=0}) 
\end{array} \right].
$$

\section{Simulation} \label{sec:simulation}
\subsection{Simulation settings}\label{ssec:sim_settings}
We conducted simulation studies to demonstrate the statistical properties of the proposed direct estimation methods for all four pattern-mixture models and to compare their performance with corresponding analogous multiple imputation-based methods.
All simulation studies share some common simulation settings which are similar to those in \cite{zhang2020likelihood} to enable a fair comparison with previously published results. Specifically, we considered a longitudinal data framework with two arms (placebo and experimental treatment), and four post-baseline visits, and 100 participants in each arm.
Thus, following the notations in Section~\ref{sec:methods}, we have $I=1$, $K=4$, and $n=200$. 

The outcome vector (including baseline value) for participants in the placebo arm was independently sampled from a multivariate normal distribution with the marginal mean vector $\bd{\mu}_0 = (0, 1.0, 1.8, 2.5, 3)'$, the marginal standard deviation (SD) vector $\bd{\sigma} = (2.0, 1.8, 2.0, 2.1, 2.2)'$, and a within-subject correlation matrix
\begin{equation}\nonumber
\bd{\rho} = 
\begin{pmatrix}
    1   & 0.6 & 0.3 & 0.2 & 0.1 \\
    0.6 & 1   & 0.7 & 0.5 & 0.2 \\
    0.3 & 0.7 & 1   & 0.6 & 0.4 \\
    0.2 & 0.5 & 0.6 & 1   & 0.5 \\
    0.1 & 0.2 & 0.4 & 0.5 & 1
\end{pmatrix}.
\end{equation}
For participants in the experimental treatment arm, assuming a differential treatment effect (alternative scenario), the marginal mean vector was set to $\bd{\mu}_1 = (0, 1.3, 2.3, 3.2, 4)'$ while the SD and the correlation matrix was assumed to be the same as for the placebo arm. Under the assumption of no treatment effect (null scenario), 
the outcome vector for subjects in the experimental treatment arm was also sampled assuming the same mean vector as for the placebo arm.

\renewcommand{\baselinestretch}{1.0}
\begin{table}[h!tb] \centering
\caption{Simulation settings and missing data proportions at the last visit based on 10000 simulated datasets}
\begin{tabular}{ccccccc}
\hline\hline
{Method} & {Setting} & {Group} & {Missing (\%)} & {Adherent (\%)} & {Parameter} & {Value}   \\ 
\hline 
\multirow{8}{1cm}{R2B, J2R, and PW}
& \multirow{4}{3cm}{No treatment effect}
& \multirow{2}{0.5cm}{P}
& \multirow{2}{1cm}{24.7}
& \multirow{2}{0.5cm}{/}
& {$\mu$} & (0, 1.0, 1.8, 2.5, 3) \\
&&&&& {$\gamma$} & (3.0, -0.2) \\
\cline{4-7}
&& \multirow{2}{0.5cm}{E}
& \multirow{2}{1cm}{24.7}
& \multirow{2}{0.5cm}{/}
& {$\mu$} & (0, 1.0, 1.8, 2.5, 3) \\
&&&&& {$\gamma$} & (3.0, -0.2) \\
\cline{2-7}
& \multirow{4}{3cm}{Differential treatment effect}
& \multirow{2}{0.5cm}{P}
& \multirow{2}{1cm}{20.1}
& \multirow{2}{0.5cm}{/}
& {$\mu$} & (0, 1.0, 1.8, 2.5, 3) \\
&&&&& {$\gamma$} & (3.2, -0.2) \\
\cline{4-7}
&& \multirow{2}{0.5cm}{E}
& \multirow{2}{1cm}{29.9}
& \multirow{2}{0.5cm}{/}
& {$\mu$} & (0, 1.3, 2.3, 3.2, 4) \\
&&&&& {$\gamma$} & (2.8, -0.2) \\
\hline
\multirow{8}{1cm}{RD}
& \multirow{4}{3cm}{No treatment effect}
& \multirow{2}{0.5cm}{P}
& \multirow{2}{1cm}{4.9}
& \multirow{2}{0.5cm}{90.1}
& {$\mu$} & (0, 1.0, 1.8, 2.5, 3) \\
&&&&& {$\eta$} & (4.0, -0.2) \\
\cline{4-7}
&& \multirow{2}{0.5cm}{E}
& \multirow{2}{1cm}{4.8}
& \multirow{2}{0.5cm}{90.2}
& {$\mu$} & (0, 1.0, 1.8, 2.5, 3) \\
&&&&& {$\eta$} & (4.0, -0.2) \\
\cline{2-7}
& \multirow{4}{3cm}{Differential treatment effect}
& \multirow{2}{0.5cm}{P}
& \multirow{2}{1cm}{4.9}
& \multirow{2}{0.5cm}{90.1}
& {$\mu$} & (0, 1.0, 1.8, 2.5, 3) \\
&&&&& {$\eta$} & (4.0, -0.2) \\
\cline{4-7}
&& \multirow{2}{0.5cm}{E}
& \multirow{2}{1cm}{10.2}
& \multirow{2}{0.5cm}{84.8}
& {$\mu$} & (0, 1.3, 2.3, 3.2, 4) \\
&&&&& {$\eta$} & (3.6, -0.2) \\
\hline
\hline 
\end{tabular}\\
\label{tab:sim_param}
\end{table}
\renewcommand{\baselinestretch}{1.5}

In the datasets for the R2B and J2R pattern-mixture model, we assumed the missing pattern to be monotone and that all baseline outcomes were observed.
Thus, $R_{jk}(i) = 0$ if $R_{j,k-1}(i) = 0$, $k = 2,\ldots,4$.
Otherwise, if $R_{j,k-1}(i) = 1$, we generated $R_{jk}(i)$ from a Bernoulli distribution with probability $\pi_{jk}(i)$ following a logistic model:
\begin{equation} \nonumber
\label{eq:sim_miss}
    \text{logit}(\pi_{jk}(i)) = \gamma_{1i} + \gamma_{2i}Y_{j,k-1}(i).
\end{equation}
Table~\ref{tab:sim_param} details the various simulation settings such as the proportion of subjects adherent to treatment, mean vectors, and the  proportion of participants with missing values for all four methods.  Values of $\gamma$, $\eta$ which control the proportion of subjects with missing values, were assumed to be the same for both placebo and experimental treatment arms under the null scenario while they were assumed to be different under the alternative scenario of differential treatment effect.


The data generation for evaluating the RD pattern-mixture model was different from those for evaluating the RTB and J2R pattern-mixture models in that the adherence (or pattern) indicator $A_{jk}(i)$ was generated prior to the missing indicator $R_{jk}(i)$.
Specifically, we assumed the adherence pattern to be monotone, and that $A_{j0}(i) = 1$ for all participants.
Thus, if the adherence indicator $A_{j,k-1}(i) = 0$, $k = 1,\ldots,4$, then $A_{jk}(i) = 0$.
Otherwise, if $A_{j,k-1}(i) = 1$, we sampled $A_{jk}(i)$ from a Bernoulli distribution with probability $\phi_{jk}(i)$ following a logistic model:
\begin{equation} \nonumber
    \text{logit}(\phi_{jk}(i)) = \eta_{1i} + \eta_{2i}Y_{j,k-1}(i,i).
\end{equation}
We explored two scenarios of different values of $(\eta_{1i}$, $\eta_{2i})$ as shown Table~\ref{tab:sim_param}.
We further assumed the monotone pattern with $R_{jk}(i) = 1$ if $A_{jk}(i) = 1$ and $R_{jk}(i) \sim \text{Bernoulli}(0.5)$ if $A_{jk}(i) = 0$.
When $R_{jk}(i)$ and $A_{jk}(i)$ were first generated in data simulation process, if the number of RDs ($a$) was greater than 5, we resampled 5 RDs from these $a$ samples and changed $R_{jk}(i)$ to 0 for the remaining $a-5$ samples. On the other hand, when $a<5$, denoting the number of samples with $R_{jk}(i)=0$ and $A_{jk}(i)=0$ to $b$, we performed resampling by two cases: (1) if $a+b \geq 5$, we resampled $5-a$ samples from these $b$ samples and set new samples to $R_{jk}(i)=1$; (2) if $a+b<5$, we set those $b$ samples to $R_{jk}(i)=1$, and also we resampled $5-a-b$ samples from adherers and set them to $A_{jk}(i)=0$.
With these settings, the number of RDs was fixed to 5, which allowed us to evaluate the model performance when there was a low number of RDs.
Lastly, we assumed a 50\% deterioration of the response for non-adherers at the last visit, that is, $Y_{j4}(i,-1) = Y_{j4}(i,i)/2$.

We independently generated 10,000 simulated samples for each scenario, and each sample was imputed 200 times when multiple imputation-based methods were applied.
In general, we compared the proposed approach to multiple imputation-based methods (using the Rubin method for the variance estimation) for each of the pattern-mixture models. The multiple imputation for RB adopted the procedures proposed by \cite{qu2022return} [Equation (6)].
Setting all intermediate measurements to missing for participants in the treatment arm with missing outcomes, the PW imputation utilized the baseline outcomes of all participants and the observed final outcome in the placebo arm to impute the missing outcomes in the treatment arm.
We performed the RD imputation within each treatment arm using the RD measurements to impute missing outcomes. All imputations under the MAR assumption were conducted by using the \textit{mlmi} and \textit{mice} packages in R. Additionally, we included the direct estimation in \cite{zhang2020likelihood} for R2B, making a fair comparison to it. 
For the newly proposed direction estimation, the analysis model to estimate $\bd{\beta}$ in Equation~\eqref{eq:psi_mar} and \eqref{eq:psi_betaplus_mar} consisted of visit $t_j$ as a factor, $\bd{X}_{j}$ as a baseline covariate, and the interaction between $t_j$ and $\bd{X}_{j}$. Besides $t_j$ and $\bd{X}_j$, the analysis model for the direct estimation in \cite{zhang2020likelihood} used treatment group $T_j$ as a factor and also modeled the outcome through the interaction between $t_j$ and $T_j$. For multiple imputation-based methods, we fitted an analysis of covariance model on the imputed datasets with $T_j$ as a factor and $\bd {X}_j$ as a covariate, while no interaction was considered in this case. The point estimate of treatment effect was the average of $10,000 \times 200$ estimates and the variance estimate was calculated by the Rubin rule. Here $\bd{X}_j$ only has one variable, which is $Y_{j0}$, the baseline value of the response variable.  
To evaluate the performance of all methods, we calculated the true mean for change from baseline with different imputation methods based on Equations~\eqref{eq:estimator_r2b}, \eqref{eq:estimator_j2r}, \eqref{eq:estimator_pw}, and \eqref{eq:estimator_rd}.

\subsection{Simulation results}\label{ssec:sim_results}

\renewcommand{\baselinestretch}{1.0}
\begin{table}[h!tb] \centering
\caption{Summary of the simulation results for the performance of the R2B pattern mixture model (based on 10000 simulated samples)}
\begin{tabular}{cccccccc}
\hline\hline
{Setting} & {Method} & {Group} & True & Bias & SD & SE & CP  \\ 
\hline \multirow{9}{4cm}{No treatment effect} & \multirow{3}{3cm}{Multiple imputation}
 & P	&	2.287   &0.002  &0.246  &0.264  &0.962\\ 
&& E	&   2.289   &0.002  &0.247  &0.264  &0.962\\ 
&& E-P	&   0.001   &0.000  &0.319  &0.374  &0.976\\ \cline{3-8} & \multirow{3}{3cm}{Direct estimation (Zhang et al)}
 & P	&	2.287   &0.002  &0.246  &0.227  &0.927\\ 
&& E	&	2.289   &0.002  &0.246  &0.227  &0.926\\ 
&& E-P	&	0.001   &0.000  &0.317  &0.321  &0.953\\ \cline{3-8}  & \multirow{3}{3cm}{Direct estimation (new)}
 & P	&	2.287   &0.003  &0.246  &0.242  &0.943\\ 
&& E	&	2.289   &0.003  &0.247  &0.243  &0.941\\ 
&& E-P  & 	0.001   &0.000  &0.319  &0.310  &0.942\\ \hline
\multirow{9}{4cm}{Differential treatment effect} & \multirow{3}{3cm}{Multiple imputation}
 & P	&	2.398   &0.001  &0.248  &0.275  &0.969\\ 
&& E	&   2.805   &0.001  &0.268  &0.279  &0.957\\ 
&& E-P	&   0.408   &0.000  &0.334  &0.393  &0.979\\ \cline{3-8} & \multirow{3}{3cm}{Direct estimation (Zhang et al)}
 & P	&	2.398   &0.001  &0.247  &0.227  &0.927\\ 
&& E	&	2.805   &0.001  &0.267  &0.255  &0.936\\ 
&& E-P	&	0.408   &0.000  &0.332  &0.341  &0.955\\ \cline{3-8}  & \multirow{3}{3cm}{Direct estimation (new)}
 & P	&	2.398   &0.003  &0.248  &0.245  &0.944\\ 
&& E	&	2.805   &0.002  &0.268  &0.265  &0.945\\ 
&& E-P  & 	0.408   &0.001  &0.333  &0.330  &0.946\\ 
\hline
\hline 
\end{tabular}\\
    {\begin{flushleft} Abbreviations: CP, coverage probability of the 95\% confidence interval; E, experimental treatment arm; P, placebo arm; R2B, return-to-baseline, SD, standard deviation of estimates of the mean; SE, mean standard error estimates of the mean.
    \end{flushleft} }
\label{tab:simu_RTB}
\end{table}

\begin{table}[h!tb] \centering
\caption{Summary of the simulation results for the performance of the J2R pattern-mixture model (based on 10000 simulated samples)}
\begin{tabular}{cccccccc}
\hline\hline
{Setting} & {Method} & {Group} & True & Bias & SD & SE & CP  \\ 
\hline \multirow{6}{4cm}{No treatment effect} & \multirow{3}{3cm}{Multiple imputation}
 & P	&	3   &0.003  &0.260  &0.176  &0.816\\ 
&& E	&   3   &0.001  &0.238  &0.168  &0.831\\ 
&& E-P	&   0   &-0.002  &0.192  &0.236  &0.983\\ \cline{3-8} &  \multirow{3}{3cm}{Direct estimation (new)}
 & P	&	3   &0.004  &0.198  &0.195  &0.944\\ 
&& E	&	3   &0.002  &0.169  &0.165  &0.944\\ 
&& E-P  & 	0   &-0.003 &0.192  &0.187  &0.945\\ \hline
\multirow{6}{4cm}{Differential treatment effect} & \multirow{3}{3cm}{Multiple imputation}
 & P	&	3   &0.000  &0.362  &0.246  &0.815\\ 
&& E	&   3.701   &0.000 &0.333  &0.244  &0.850\\ 
&& E-P	&   0.701   &0.000 &0.253  &0.337  &0.992\\ \cline{3-8} &  \multirow{3}{3cm}{Direct estimation (new)}
 & P	&	3   &0.003  &0.275  &0.270  &0.944\\ 
&& E	&   3.701   &0.003  &0.235  &0.234  &0.947\\ 
&& E-P	&   0.701   &0.000  &0.253  &0.247  &0.944\\ 
\hline
\hline 
\end{tabular}\\
    {\begin{flushleft} Abbreviations: CP, coverage probability of the 95\% confidence interval; E, experimental treatment arm; J2R, jump-to-reference; P, placebo arm; SD, standard deviation of estimates of the mean; SE, mean standard error estimates of the mean.
    \end{flushleft} }
\label{tab:simu_J2R}
\end{table}

\begin{table}[h!tb] \centering
\caption{Summary of the simulation results for the performance of the PW patterm-mixture model (based on 10000 simulated samples)}
\begin{tabular}{cccccccc}
\hline\hline
{Setting} & {Method} & {Group} & True & Bias & SD & SE & CP  \\ 
\hline \multirow{6}{4cm}{No treatment effect} & \multirow{3}{3cm}{Multiple imputation}
 & P	&	3   &0.003  &0.198  &0.175  &0.918\\ 
&& E	&   2.952   &0.000  &0.170  &0.178  &0.959\\ 
&& E-P	&   -0.048   &-0.003  &0.193  &0.244  &0.986\\ \cline{3-8} &  \multirow{3}{3cm}{Direct estimation (new)}
 & P	&	3   &-0.005  &0.198  &0.195  &0.945\\ 
&& E	&	2.952   &0.000  &0.170  &0.168  &0.945\\ 
&& E-P  & 	-0.048   &-0.005  &0.193  &0.189  &0.944\\ \hline
\multirow{6}{4cm}{Differential treatment effect} & \multirow{3}{3cm}{Multiple imputation}
 & P	&	3   &0.002  &0.274  &0.246  &0.918\\ 
&& E	&   3.642   &0.001  &0.239  &0.266  &0.968\\ 
&& E-P	&   0.642   &-0.001  &0.255  &0.353  &0.992\\ \cline{3-8} &  \multirow{3}{3cm}{Direct estimation (new)}
 & P	&	3   &0.003  &0.275  &0.271  &0.945\\ 
&& E	&   3.642   &0.001  &0.239  &0.244  &0.953\\ 
&& E-P	&   0.642   &-0.002  &0.255  &0.255  &0.950\\ 
\hline
\hline 
\end{tabular}\\
    {\begin{flushleft} Abbreviations: CP, coverage probability of the 95\% confidence interval; E, experimental treatment arm; P, placebo arm; PW, placebo washout; SD, standard deviation of estimates of the mean; SE, mean standard error estimates of the mean.
    \end{flushleft} }
\label{tab:simu_PW}
\end{table}

\begin{table}[h!tb] \centering
\caption{Summary of the simulation results for the performance of the RD pattern-mixture model (based on 10000 simulated samples)}
\begin{tabular}{cccccccc}
\hline\hline
{Setting} & {Method} & {Group} & True & Bias & SD & SE & CP  \\ 
\hline \multirow{6}{4cm}{No treatment effect} & \multirow{3}{3cm}{Multiple imputation}
 & P	&	2.852   &0.002  &0.253  &0.217  &0.903\\ 
&& E	&   2.852   &0.003  &0.254  &0.217  &0.904\\ 
&& E-P	&   0.001   &0.001  &0.308  &0.307  &0.949\\ \cline{3-8} &  \multirow{3}{3cm}{Direct estimation (new)}
 & P	&	2.852   &0.002  &0.257  &0.252  &0.946\\ 
&& E	&	2.852   &0.004  &0.256  &0.252  &0.943\\ 
&& E-P  & 	0.001   &0.002  &0.312  &0.307  &0.948\\ \hline
\multirow{6}{4cm}{Differential treatment effect} & \multirow{3}{3cm}{Multiple imputation}
 & P	&	2.852   &0.002  &0.253  &0.218  &0.905\\ 
&& E	&   3.695   &0.003  &0.261  &0.219  &0.903\\ 
&& E-P	&   0.844   &0.001 &0.313  &0.310  &0.949\\ \cline{3-8} &  \multirow{3}{3cm}{Direct estimation (new)}
 & P	&	2.852   &0.002  &0.257  &0.252  &0.946\\ 
&& E	&   3.695   &0.003  &0.265  &0.259  &0.942\\ 
&& E-P	&   0.844   &0.001  &0.318  &0.313  &0.948\\ 
\hline
\hline 
\end{tabular}\\
    {\begin{flushleft} Abbreviations: CP, coverage probability of the 95\% confidence interval; E, experimental treatment arm; P, placebo arm; RD, retrieved dropout; D, standard deviation of estimates of the mean; SE, mean standard error estimates of the mean.
    \end{flushleft} }
\label{tab:simu_RD}
\end{table}
\renewcommand{\baselinestretch}{1.5}

In this section, we discuss the results of simulation studies as introduced in Section~\ref{ssec:sim_settings}.
In Tables~\ref{tab:simu_RTB} through \ref{tab:simu_RD}, the first column in the summary statistics part of the table represents the true value of $\mu_{04}^\text{MAR}$, $\mu_{14}^\text{MAR}$, and $\mu_{14}^\text{MAR} - \mu_{04}^\text{MAR}$ for each imputation method across 10000 simulated datasets.
The next four columns of the summary statistics list the average bias between estimates and true values, the SD of the estimates, the mean standard error (SE) estimates, and the coverage probability of the 95\% confidence interval (CI) for the mean change from baseline at each treatment arm and treatment difference based on 10000 simulations.

Table~\ref{tab:simu_RTB} shows the results for the R2B pattern-mixture model from three estimation methods based on multiple imputation, the direct estimation by \cite{zhang2020likelihood}, and the direct method proposed in this article. The biases were very similar between the three methods, but the standard errors were different. The SEs for the methods based on multiple imputation were larger than the corresponding SDs of the mean estimates, and the corresponding coverage probabilities of the 95\% CIs were larger than 0.95, suggesting the Rubin method overestimates the variance. \cite{zhang2020likelihood}'s method underestimates the standard error for each treatment group, resulting in lower coverage probability of the 95\% CI. The newly proposed direct estimation method produced SE estimates close to the observed SDs of the mean estimates and 95\% CIs with approximate 0.95 coverage probabilities. {The coverage probability for the direct estimation method was slightly lower than 0.95 (as low as 0.941). It was likely because the asymptotic inference was not ideal for a total sample size of 200 (100 per arm). Additional simulation with increased sample size improved the coverage probability to between 0.946 and 0.951 for a total sample size of 1,000.} 

Tables 3, 4, and 5 show the simulation results for the J2R, PW, and RD pattern-mixture models, respectively. For all three pattern-mixture models, the proposed direct estimation methods and methods based on multiple imputation all produced estimates with little bias, but only the proposed direct estimation methods provided approximately consistent SEs, and 95\% CIs with approximately correct coverage probabilities. 

\section{Application} \label{sec:application}
In this section, we applied the proposed direct estimation approaches to data from a clinical trial of participants with type 2 diabetes (AWARD-1, ClinicalTrials.gov Identifier: NCT01064687). The AWARD-1 study was a randomized placebo- and active-controlled trial evaluating the efficacy of two doses of dulaglutide versus exenatide and placebo on the ability to lower HbA1c in participants with type 2 diabetes. The study was conducted in accordance with the Declaration of Helsinki guidelines on good clinical practices \citep{world2010world}. A total of 978 patients were randomized to dulagludite 0.75 mg, dulaglutide 1.5 mg, exenatide 10 $\mu$g, and placebo in a 2:2:2:1 ratio. The primary outcome was the mean change in HbA1c from baseline to 26 weeks. In this article, we only considered the treatment groups of placebo, dulaglutide 0.75 mg, and dulaglutide 1.5 mg. 
In this post hoc analysis, the estimand of interest was the treatment difference between each dulaglutide dose and placebo in the mean change in HbA1c from baseline to 26 weeks among all patients who underwent randomization, regardless of treatment adherence or using ancillary therapies. Treatment discontinuations and the use of rescue medications were handled by the treatment policy strategy. 

\renewcommand{\baselinestretch}{1}
\begin{table}\centering
\caption{Summary of treatment discontinuations and missing data at Week 26 in the AWARD-1 study}
\begin{tabular}{lccc}
\hline\hline
\multirow{2}{6cm}{Description} & Placebo & Dulaglutide 0.75 mg & Dulaglutide 1.5 mg \\
 & (N=141) & (N=280) & (N=279) \\
\hline 
Adhered to treatment with \\ $\;\;\;$ 
non-missing primary outcome & 122 (86.5\%) & 255 (91.1\%) & 254 (91.0\%) \\
Adhered to treatment but with \\ $\;\;\;$  missing primary outcome data & 2 (1.4\%) & 1 (0.36\%) & 0 (0.0\%) \\
Discontinued treatment with \\ $\;\;\;$ non-missing primary outcome \\ $\;\;\;$ (retrieved dropouts) & 5 (3.5\%) & 9 (3.2\%) & 9 (3.2\%) \\
Discontinued treatment with \\ $\;\;\;$ missing primary outcome & 12 (8.5\%) & 15 (5.4\%) & 16 (5.7\%) \\
\hline\hline
\end{tabular}
\label{tab:award1desc}
\end{table}
\renewcommand{\baselinestretch}{1.5}


\renewcommand{\baselinestretch}{1.0}
\begin{table}[h!tb] \centering
\caption{Summary of mean estimates (SE) of change from baseline to Week 26 in HbA1c using direct estimation  approaches to data from AWARD-1 Study.}
\begin{tabular}{{p{2cm}p{2cm}p{2cm}p{2cm}p{2cm}p{2cm}}}
\hline\hline
{Method} & {Placebo} & {Dula 0.75 mg} & Dula 1.5 mg & Dula 0.75 mg vs. Placebo & Dula 1.5 mg vs. Placebo \\ 
\hline 
R2B  & -0.59 (0.08)	&	-1.24 (0.06)   & -1.46 (0.05)  & -0.65 (0.09)  & -0.87 (0.09) \\ 
{R2B (Zhang)} & {-0.58 (0.06)} & {-1.24 (0.05)} & {-1.46 (0.05)} & {-0.67 (0.08)} & {-0.89 (0.08)} \\
J2R & -0.66 (0.09) & -1.28 (0.05) & -1.50 (0.05) & -0.62 (0.09) & -0.84 (0.09) \\
PW &  -0.66 (0.09) & -1.29 (0.05) & -1.50 (0.05) & -0.63 (0.09) & -0.84 (0.09) \\
RD & -0.66 (0.08) & -1.30 (0.06) & -1.53 (0.05) & -0.65 (0.09) & -0.87 (0.09) \\
\hline
\hline 
\end{tabular}\\
    {\begin{flushleft} Abbreviations: Dula 0.75 mg, Dulaglutide 0.75 mg arm; Dula 1.5 mg, Dulaglutide 1.5 mg arm; SE, standard error; R2B, Return to baseline; J2R, Jump to reference; PW, Placebo washout; RD, Retrieved dropouts.
    \end{flushleft} }
\label{tab:application_result}
\end{table}
\renewcommand{\baselinestretch}{1.5}


\begin{figure}
\centering
\includegraphics[scale=0.7]{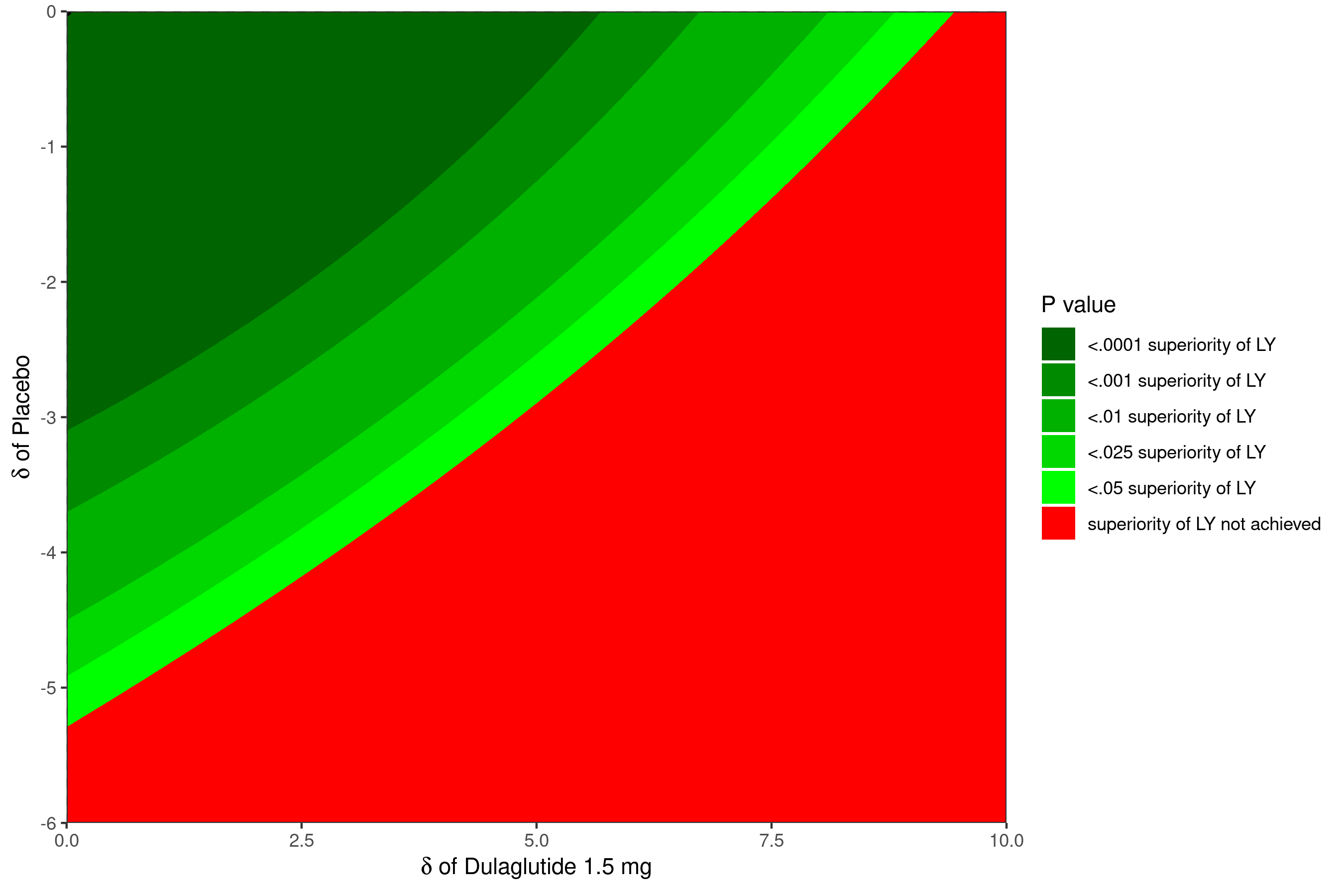}
\caption{Plot of two-way tipping point boundaries for various significance thresholds comparing the dulaglutide 1.5 mg with placebo using the retrieved dropout estimation method. The red region indicates the area where superiority of dulaglutide 1.5 mg to placebo would not be achieved (i.e., $p> 0.05$).}
\label{fig:rd_tip}
\end{figure}

Table~\ref{tab:award1desc} summarizes the treatment and study discontinuation and the proportion of patients with missing primary outcome measurement. Since very few participants who adhered to treatment had missing values, all participants with a missing primary outcome measurement were considered to belong to Pattern B in the direction estimation procedure for each of the pattern-mixture models: R2B, J2R, PW, and RD. Overall, 14 (10\%), 16 (6\%), and 16 (6\%) of participants had missing data for the primary outcome in placebo, the 0.75-mg dulaglutide, and the 1.5mg-dulaglutide  groups, respectively.

Table~\ref{tab:application_result} shows the estimates for the primary outcome for each treatment group and the differences relative to placebo after applying the direct estimation methods to the AWARD-1 study data for the aforementioned estimands. Under all four patterns employed for missing data, the treatment effects for both dulaglutide arms relative to placebo were highly significant. {The R2B estimates using the methods described in this article and by  \cite{zhang2020likelihood} produced very similar results. In overall,} the R2B estimates had the least mean reduction from baseline for each of the three treatment arms compared with the other three approaches. 
The placebo arm on an average showed a good response in terms of mean change in HbA1c from baseline to 26 weeks. This is likely because of the residual effect of the up-titration of metformin and pioglitazone during the lead-in period.  Therefore, the J2R and PW methods, which impute the missing values using placebo data, led to better estimates for HbA1c reduction from baseline for all three treatment groups compared to the R2B method. Since there might be some residual effect of the treatment for subjects with premature treatment discontinuation, the RD method also led to better HbA1c reduction from baseline compared to the R2B method. 
Among all four pattern-mixture models, RD had the largest estimates for the treatment effect, which was likely because the RD outcomes  continued to be influenced by the treatment effect realized before treatment discontinuation. The estimated mean differences (SE) in HbA1c relative to placebo with the RD imputation were -0.65 (0.09) and -0.87 (0.09) for the dulaglutide 0.75-mg and 1.5-mg groups, respectively.

The two-way $\delta$-adjusted and tipping point analyses for the AWARD-1 study data were also performed using the methods described in Section~\ref{sec:delta_adjusted}. The results from comparing dulaglutide 1.5 mg with placebo for the direct estimation using the RD pattern are shown in Figure~\ref{fig:rd_tip}. The $x$-axis represents the $\delta$ added to the dulaglutide 1.5-mg group, while the $y$-axis shows the $\delta$ added to the placebo group. The study result from this comparison, without addition of penalty values to both the treatment groups, corresponds to the $(0,0)$ point on the plot, which is highly significant indicated by the dark green region ($p<0.0001$). The different colored regions represent the tipping point boundaries, which are denoted by different p-value intervals that become less significant with each color, culminating in the red region that corresponds to a non-significant treatment comparison ($p>0.05$).
For example, because of the small proportion of participants with missing data and the large estimated treatment effect without $\delta$-adjustment, extreme penalty values such as $-5$ to the placebo arm and $+2$ to the dulaglutide 1.5-mg group would need to be added to the imputed values to make this comparison statistically non-significant. This shows that the treatment effects seen in the AWARD-1 study estimated using the RD pattern-mixture model were highly robust and not likely to be affected by the missing data seen in the study. Similar $\delta$-adjusted and tipping point analyses were performed for other imputation methods and similar results were observed (data are not shown). 

\section{Summary and discussion}
\label{sec:summary}
With the release of ICH E9 (R1) guidance for estimands and sensitivity analyses, estimands using the treatment policy strategy to handle intercurrent events have been widely used as the primary estimands in clinical trials. Therefore, it is important to use the appropriate analysis models including efficient use of imputation methods for handling missing values in estimating the corresponding estimands. \cite{wang2023statistical} discussed a few imputation methods corresponding to the treatment policy estimands. In this article, we {first clearly described the potential outcome to impute (which has not been done in literature) and then} provided corresponding analytical methods without the need for explicit imputation. There are two advantages to use these analytical approaches: (1) the computation is fast, and (2) these methods always provide consistent variance estimation with the correct coverage probability for the CI. Under traditional multiple imputation methods for handling missing data, the SE might be estimated incorrectly when using the Rubin variance estimation, resulting in incorrect coverage probability for the CI. {Bootstrap can be combined with multiple imputation to obtain the appropriate variance and confidence interval estimation \citep{bartlett2020bootstrap}, but the computation is time consuming. Additionally, the analytic approach in this article allows for the imputation for estimands using a mixed hypothetical and treatment policy strategies to handle intercurrent events. }

{In the simulation, we generated missing data assuming the probability of missingness depends on the observed outcomes. This is not equivalent to ignorable missingness or MAR \citep{rubin1976inference}. The ignorability assumption assumes the outcome for the missing values (if not missing) follows the same conditional distribution as the outcome of the observed value. In the patter mixture models we discussed in this article, the potential outcome for the missing values is assumed to deviate from the conditional distribution of the observed values and follows a special pattern, such R2B or J2R. Therefore, essentially we are dealing with non-ignorable missingness even though the probability of missingness only depends on the observed values. Of course, one could make the probability of missingness depend on an unobserved outcome in the simulation, but the proposed estimators for these pattern-mixture models would be biased. Note these pattern mixture models are rarely true in the real clinical trials, but mostly rather a conservative approximation. In this article, all we tried to achieve is to provide an analytic estimator as that based on multiple imputation, not to provide a new and more robust estimator, for the commonly used pattern-mixture models in clinical trials. In addition, this article does not provide analytic estimators for other imputation methods, such as imputation using the worst outcomes, which may be studied in future research.
}

Recently, \cite{garcia2023flexible} also developed an analytical framework for some imputation methods such as J2R and CR. In comparison, our research provides analytical methods for the R2B, PW, and RD pattern-mixture models, a systematic adjustment for baseline covariates, and an analytic approach for the $\delta$-adjusted analysis with direct identification of the tipping points. While \cite{garcia2023flexible} discussed using differential slopes for baseline covariates as an extension, the analytical methods in this article consider differential slopes as the default, which provides increased flexibility in modeling slopes between treatment groups. In addition, our proposed methods allow for mixed ``imputation" strategies with certain missing values imputed based on the MAR assumption and others imputed by one of the R2B, J2R, PW, and RD patterns. 

In summary, when the treatment policy estimand strategy {or a mixture of the treatment policy and hypothetical strategies} is used to handle intercurrent events, the proposed direct estimation {of the commonly used pattern mixture models provides} a convenient and robust approach for analyzing clinical trial data. 

\section*{Appendix. Supplemental material}
{The GitHub link \url{https://github.com/jitong-lilly/MIDIR} provides the relevant R code(s) to simulate an example dataset which is similar in summary statistics and missing information to the real life clinical trial data used in the manuscript and produce results under direct estimation methods for the simulated example. }

\section*{Acknowledgement}
All authors would like to thank Drs Jeremiah Jones, Rong Liu, and Jianghao Li for useful comments and reviewing the programs.

\section*{Disclosure statement}
All authors are employees and minor shareholder of Eli Lilly and Company.

\bibliographystyle{apalike}
\bibliography{references}

\end{document}